 \documentclass[final,5p,times,twocolumn]{elsarticle}

\usepackage{float}

\usepackage{graphicx}
 \usepackage{epsfig}
\begin{document}
\def\ds{\displaystyle}
\begin{frontmatter}

\title{Linear Rogue waves}
\author{C. Yuce }
\address{ Physics Department, Anadolu University, Eskisehir, Turkey}
\ead{cyuce@eskisehir.edu.tr} \fntext[label2]{}
\begin{abstract}
We predict the existence of linear discrete rogue waves governed by the discrete nonlinear Schrodinger equation. We discuss that Josephson effect is the underlying reason for the formation of such waves. 
\end{abstract}


\end{frontmatter}


\section{Introduction}

Rogue waves, sometimes known as freak waves or extreme waves, are doubly localized waves in both space and time and appear as a peak on a finite background. They have been rarely observed in oceans and but recently realized in a number of different physical contexts such as optics \cite{opt2}, microwaves \cite{linear1} and inhomogeneous media \cite{linear2}. Rogue wave solutions have been studied mostly for the focusing nonlinear Schrodinger equation (NLS). Three exact analytical solutions of the NLS, Peregrine soliton \cite{pereg}, Kuznetsov and Ma soliton \cite{kuz,ma} and Akhmediev breather \cite{akhe}, clearly explains the dynamics of rogue waves. In the Peregrine soliton, a slight modulation on a uniform background grows until it reaches its maximum value as a result of the modulational instability and then decays and vanishes. The Peregrine soliton was experimentally realized in an optical system in 2010 \cite{deney} and then in a water wave tank in 2011 \cite{deney2}. The Peregrine soliton is the first order rational solution of the NLS and the second-order rational solution was studied in \cite{takivv} and observed experimentally in \cite{super}. The ratio of maximum amplitude of the rogue wave to the background amplitude is $3$ for the first order rational solution while it is $5$ for the second order one. The extension of rogue waves to discrete systems has been investigated by some authors \cite{AL1,AL2,AL3,AL4,AL5,AL6,AL7,AL8,extreme1,extreme2}. The key idea to construct a discrete rogue wave is to use a proper choice of initial field amplitude.\\
As it is noted in a recent review article \cite{reviewpap}, the originating mechanism of rogue waves is still a matter of debate (references therein). The modulational instability is the most popular one but is restricted to nonlinear systems. The interpretation of rogue waves in terms of the two Akhmediev breather collisions was also made for the nonlinear system \cite{coll}. We would like to emphasize that rogue waves can occur even in the absence of nonlinear interaction. Rogue waves has recently been observed in a linear microwave system \cite{linear1}. The authors in \cite{linear1} discussed that random positions of metal cones acts like a "bad lens" directing waves towards some focal points and the intensity becomes five or more times greater than background level. In  \cite{linear2}, rogue waves in a linear inhomogeneous media were analyzed and it was discussed that granularity and spatial inhomogeneity are the two joint generators of optical rogue waves. In \cite{linear0}, linear rogue wave formation in oceans was analyzed. A uniformly distributed initial currents is supposed to enter the 2D spatial system along the $y$ axis and propagate along $x$-axis. The currents are considered to have a single velocity but a Gaussian angular spread ( the components of the initial velocity in the $x$ and $y$ directions change with $y$). It was shown that the currents focus along the propagation direction and consequently the density becomes high enough to be interpreted as a rogue event. The focusing effect was also analyzed in \cite{linear3}. \\
Motivated by the investigations of rogue waves in linear systems and by the question of underlying mechanism for the appearance of large amplitude waves, we inquire whether rogue waves occur in linear domain for a discrete system. In this paper, we show the existence of linear discrete rogue waves governed by the discrete nonlinear Schrodinger equation and discuss the underlying mechanism for such waves.

\section{Linear Discrete Rogue Waves}

The propagation of an optical field in a tight binding waveguide array or time evolution of matter wave in a tight binding  optical lattice can be described by the discrete nonlinear Schrodinger equation (DNLS). It is a nonintegrable equation and posseses two conserved quantities, the total number of particles (or power) and the total energy. We consider DNLS to study formation and evolution of discrete rogue wave.
\begin{equation}\label{temel}
i\frac{d\Psi_n}{dt}=-J(\Psi_{n+1}+\Psi_{n-1})+g|\Psi|_n^{2}\Psi_n
\end{equation}
where $\Psi_n$ is the complex wave function at the $n$-th waveguide, $t$ is time ($t$ is replaced by the propagation direction $z$ in an array of waveguide), $J$ is the coupling coefficient between $n$-th waveguide and adjacent waveguides, $g$ is the nonlinear interaction strength. Our aim is to find discrete rogue wave solution of the DNLS not triggered by modulational instability. We will firstly consider the non-interacting limit $g=0$ and then investigate the effect of nonlinear interaction.  \\
Generally speaking, a discrete rogue wave is the strong localization over a few lattice sites. In \cite{konop}, it was discussed that initial conditions must be chosen properly to observe discrete rogue waves. The subsequent dynamics is strongly sensitive to initial wave function. We look for an initial wave function that leads to rogue waves even in the absence of nonlinear interaction. We propose to start with the following uniform wave function
\begin{equation}\label{ic00}
\Psi_n(t=0)=A~ e^{i \Phi(n)}~,~~~~\Phi(n)=\sum_{i=1}^{\infty} c_i (n-n_i)^i  
\end{equation}
where the background amplitude $\ds{A}$ is supposed to be a real number, the coefficients $\ds{c_i}$ and the shifts $\ds{n_i}$ are constant and the initial phase $\Phi(n)$ is position dependent. The initial density, $\ds{|\Psi(t=0)|^2}$, is uniform along the lattice and this wave function is therefore a good candidate to look for rogue waves that appear from nowhere. The coefficients $\ds{c_i}$ are undetermined. There are infinitely many terms in the expression of the phase $\Phi(n)$. Fortunately, principal aspects of linear rogue waves are captured by assuming that only two of them are non-vanishing: $\ds{c_1\neq0}$, $c_2\neq0$ and $c_i=0$ for $i>2$. As a special case, the wave function becomes an exact solution when $c_2=0$ and $c_1\neq0$. This solution is known as modulationially unstable. If $c_2\neq0$, the time evolution changes significantly as we shall see below. No exact solution is available even in the absence of nonlinear interaction, we therefore perform numerical computation to find the time evolution of the above initial wave function. In our numerical computation, we solve the equation (1) for the periodical boundary condition, $\ds{\Psi_{N+1}(t)=\Psi_1(t)}$, where $N$ is the total number of lattice sites. Suppose first the noninteracting limit, $g=0$ and take $J=1$, $A=0.4$ and $N=200$. The Fig-1 plots the time evolution of the above initial wave for three different parameters. These figures clearly show that time evolution is sensitive to the exact numerical values of $c_1$ and $c_2$. The Fig-1.a, Fig-1.b and Fig-1.c are for $c_1=\pi$, $c_2= 0.005 $ (single peak), $c_1=0$, $c_2=0.03$ (W-shaped) and $c_1=0$, $c_2=0.06$ (zigzag-shaped), respectively. In all the of figures, we see some peaks that show strong localizations of particles. More precisely, particles are focused towards some focal points and then defocused like light in an imperfect lens in optics. The heights of the peaks are nearly three times larger than the background amplitude. The peaks appear from nowhere and disappear shortly as a result of the tunneling of the particles to adjacent lattice sites. Therefore, they can be interpreted as discrete rogue waves. They aren't due to modulational instability since the nonlinear interaction is assumed to be zero. Let us analyze the three figures in more detail. In the Fig-1.a, we see a single rogue wave. The ratio of the absolute of the maximum amplitude to the background amplitude is equal to $3$. The particles are focused towards the center of the lattice. A single peak is formed at around $t=50$ and then decays in a short time. The decaying mechanism is in fact defocusing of focused particles due to the tunneling to adjacent sites. When the particles are defocused to all sites, the system starts to behave chaotically. (In the Fig.1, the chaotic behaviors are not shown) Let us now discuss the Fig.1.b and Fig.1.c. There are now more than one focal point and consequently more than one peak. Since the particles are distributed in the peaks, the heights of the peaks decrease and consequently the widths of the peaks increase. The latter one shows us that the lifetime of the rogue waves increases. We would like to note that rogue waves are formed earlier in the last two figures. As in the Fig.1.a, the particles in each peaks are spread mainly into two lobes and the system behaves chaotically at large times. To this end, we note that changing the background density $A$ doesn't change the rogue wave pattern since the system is linear. However varying the tunneling parameter $J$ changes the time at which the peaks appear. If the peaks occur at $t=\tau$ when $J=1$, then they would occur at $t=\tau/J_0$ when $J=J_0$. This scaling can be easily seen from the equation (1).\\
Let us now study the underlying mechanism of the discrete rogue wave formation in the linear domain. The key idea here is the Josephson effect \cite{jos}. Our initial wave function (2) is uniform but has a position dependent phase. This leads to a position dependent current in the lattice. The current in turn leads to population imbalance between adjacent lattice sites. If the position dependent phase is chosen appropriately, these currents add up altogether to form a high density peak at some points (the focal points). The Josephson current can easily be understood for the exactly solvable two-site system. We follow Feynman's simple two-state phenomenological approach \cite{feynman} to describe Josephson effect. Assume that coupling between the the two lattice sites equals $J$ and $\ds{\Psi_1}$ and $\ds{\Psi_2}$ represent wave functions on the first and second sites, respectively. We further assume that the densities at each sides are initially equal. Therefore the initial wave functions are given by $\ds{\Psi_1=\sqrt{N_0}e^{i\theta_1}}$ and $\ds{\Psi_2=\sqrt{N_0}e^{i\theta_2}}$, where $\theta_1$ and $\theta_2$ are the phases on the two sides and $N_0$ is the density of particles. This phase difference between the lattice sites induces a current, known as Josephson current, which is proportional to $ J n_0 \sin(\theta_2 -\theta_1)$ \cite{feynman}. Our system is indeed a many lattice site generalization of this two site system. The Josephson current is position dependent since the relative phase of two sites is position dependent. The induced Josephson current is symmetrical with respect to the center and maximum at the edges of the lattice when $c_2\neq0$. So, the population imbalance occurs first at the edges and particles start to flow towards the center. Under some appropriate conditions, the currents at each sites don't cancel each other and particles are focused to some points, where peaks occur. As a result, we say that modulational instability is not responsible for rogue waves in our system. Instead an analogue of a Josephson-like effect can be used to explain the formation of linear discrete rogue waves.\\
\begin{figure}[t]\label{20}
\includegraphics[width=6.0cm]{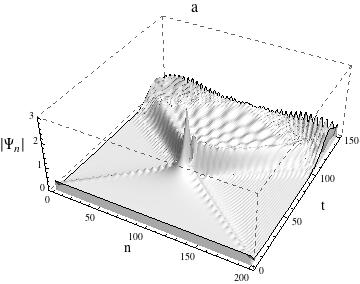}
\includegraphics[width=6.0cm]{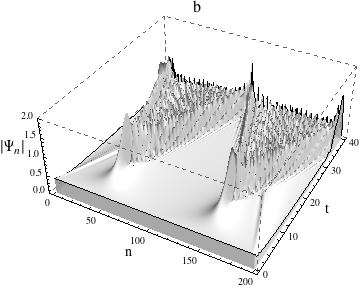}
\includegraphics[width=6.0cm]{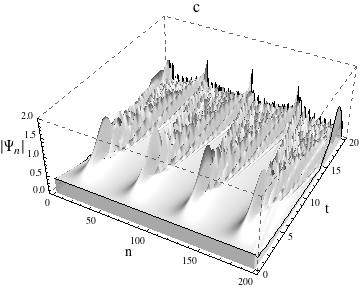}
\caption{The absolute of the wave function for $N=200$, $\ds{g=0}$,  $A=0.4$ $n_1=N/2$ and $n_2=0$. The system has a single peak when $c_1=\pi$ and $c_2=  0.005 $ (a)
W-shaped structure when $c_1=0$ and $c_2=0.03$ (b) and zigzag-shaped structure when $c_1=0$ and $c_2=0.06$ ( c).     }
\end{figure}
Let us now discuss robustness of rogue waves in the linear domain. Suppose the same parameters already used in the Fig-1.a. We consider robustness against noise on the initial density. We expect that rogue waves are robust against noise. This is because Josephson current between adjacent sites is not destroyed in the presence of such noise. The Josephson current exists even when the system is not uniformly distributed at $\ds{t=0}$. In our numerical computation, we add a large amount of noise: $\Psi_n=A(1+0.3 W_n)e^{i\Phi_n}$, where $W_n$ are random numbers with zero-mean distribution in the interval $\ds{\left[-1,1\right]}$. The Fig-2.b displays time evolution of this initial wave packet. We see some fluctuations on the background and a single peak as we expect. Let us now study robustness against diagonal disorder \cite{cemsemiha}. We expect a similar picture since random on-site potential doesn't destroy Josephson current, either. As can be seen from the Fig.2.a, the pattern formation remains almost the same even when the diagonal disorder is large.
\begin{figure}[t]\label{20}
\includegraphics[width=6.0cm]{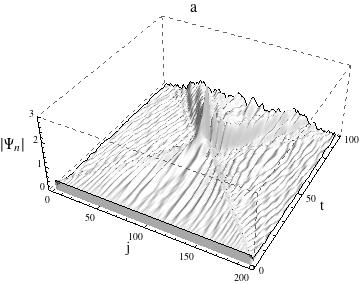}
\includegraphics[width=6.0cm]{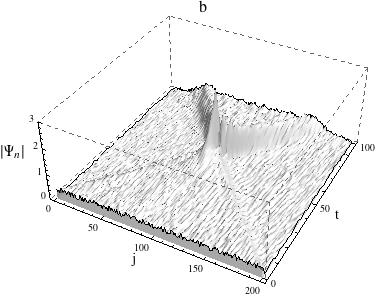}
\caption{ The same parameters as in the Fig.1.a are used.  The peak is robust against even high amount of noise as large as $5$ percentage diagonal disorder (a) and  $30$  percentage noise (b).  }
\end{figure} 

\subsection{The effect of Noninear Interaction}

Let us now consider $\ds{g\neq0}$. Positive and negative values of $\ds{g}$ correspond to an repulsion and attraction. It is well known that rogue waves can be formed on a finite background if the system is modulationally unstable. In this subsection, our aim is not to study rogue wave formation due to modulational instability. Instead, we analyze the effect of nonlinear interaction on the process already present in the linear system. \\
During the formation of the peak, the nonlinear interaction plays an important role. The nonlinear interaction leads to either self-focusing or self-defocusing. In the limiting case $c_2=0$, both self-focusing and self-defocusing occur depending on the values of $c_1$ at fixed nonlinear interaction. This was experimentally achieved in \cite{moran}. Analogously, we also expect self-focusing and self-defocusing even when $c_2\neq0$. Let us now introduce attractive nonlinear interaction to the system described in the Fig-1.a. If $c_2=0$, we would see self-defocusing since $c_1=\pi$ \cite{moran}. We also numerical see self-defocusing when $c_2=0.005$. The peak density is lowered by the attractive interaction and the width consequently increases. If attractive interaction is strong enough, then the peak is completely destroyed as can be seen in the Fig.3.a. In other words, rogue wave is not formed for strong attractive nonlinear interaction. We would like to note that rogue waves given in the Fig.1.b and Fig.1.c. are destroyed at much stronger attractve nonlinear interaction. The physics of rogue wave is drastically changed if repulsive nonlinear interaction is introduced. If $g$ is big enough, then the peak doesn't vanish because of the self-trapping mechanism as can be seen in the Fig.3.b. Chaotic fluctuations occur on the background while the rogue wave survives with a slight change on its amplitude in time. The discrete rogue wave that appears from nowhere and survives for a long time was studied in \cite{cemsemiha}. Note that self-trapping depends on $c_1$ and $c_2$. For example, we don't numerically see long-living discrete rogue waves in the presence of repulsive nonlinear interaction if the same parameters in the Fig.3.b and Fig.3.c are used. 
\begin{figure}[t]\label{220}
\includegraphics[width=6.0cm]{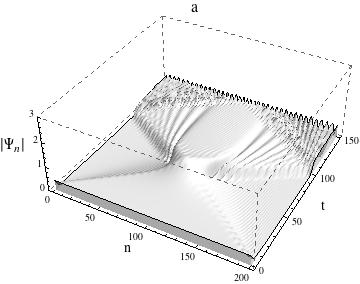}
\includegraphics[width=6.0cm]{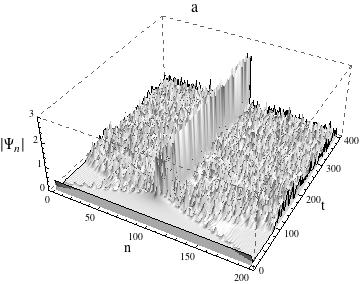}
\caption{The same parameters as in the Fig.1.a but $g=-1$ (a) and $g=1$ (b) are used. The single peak is destroyed when the attractive nonlinear interaction is strong enough and long-living discrete rogue wave is formed when repulsive interaction is strong enough. }
\end{figure}\\
To sum up, we have predicted the existence of discrete linear rogue waves. The underlying physics for the formation of linear rogue waves can not be explained by modulational instability. We have discussed that Josephson effect can be used to explain linear discrete rogue waves. It is tempting to find an exact rogue wave solution (not triggered by modulational instability) for the nonlinear Schrodinger equation. Finally, we say that an experimental realization of our predictions can be achieved with current technology.

\end{document}